# Quantifying the Role of the Surfactant and the Thermophoretic Force in Plasmonic Nano-Optical Trapping

Quanbo Jiang,[1] Benoît Rogez,[1] Jean-Benoît Claude,[1] Guillaume Baffou,[1] Jérôme Wenger[1,*]

[1] Aix Marseille Univ, CNRS, Centrale Marseille, Institut Fresnel, 13013 Marseille, France

[*] Corresponding author: jerome.wenger@fresnel.fr

**Abstract**

Plasmonic nano-tweezers use intense electric field gradients to generate optical forces able to trap nano-objects in liquids. However, part of the incident light is absorbed into the metal, and a supplementary thermophoretic force acting on the nano-object arises from the resulting temperature gradient. Plasmonic nano-tweezers thus face the challenge of disentangling the intricate contributions of the optical and thermophoretic forces. Here, we show that commonly added surfactants can unexpectedly impact the trap performance by acting on the thermophilic or thermophobic response of the nano-object. Using different surfactants in double nanohole plasmonic trapping experiments, we measure and compare the contributions of the thermophoretic and the optical forces, evidencing a trap stiffness 20× higher using sodium dodecyl sulfate (SDS) as compared to Triton X-100. This work uncovers an important mechanism in plasmonic nano-tweezers and provides guidelines to control and optimize the trap performance for different plasmonic designs.

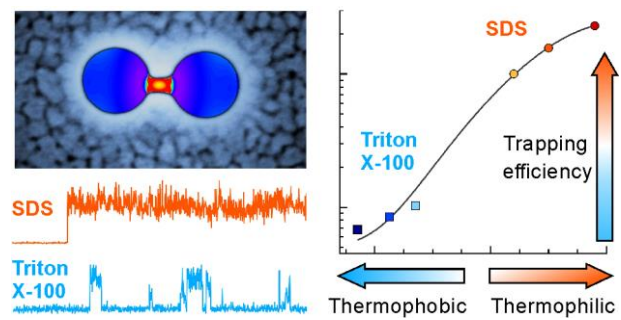

Figure for Table of Contents

Plasmonic trapping opens a powerful avenue to overcome the limitations of conventional optical tweezers.[1–6] Nanoparticles,[7–23] quantum dots,[24,25] proteins,[4,26–29] or DNA molecules[30,31] can be trapped, paving the way for applications in biochemistry,[32–34] life sciences,[35–37] and quantum information processing.[38–42] While the optical gradient force is now well understood,[43] the physics of plasmonic trapping is more complex as optical, thermal and fluidic effects are intrinsically entangled.

Plasmonics is bound to absorption losses into the metal, which induce a local temperature gradient around the nanostructure.[44–48] This temperature gradient then exerts a thermophoretic force on the nano-object as a consequence of the Ludwig-Soret effect.[49–52] The main physical principle here is that the thermal gradient generates an interfacial fluid flow at the nano-object surface, which in turn moves the object across the solution. While the thermophoretic force can be beneficial to trap some nano-objects,[51,53–58] it has also been reported to disturb the trap potential.[12,13,33,46,52,59–61] As a consequence of this apparent confusion and the lack for a proper characterization method, the influence of the thermophoretic force and its potential to improve the trap performance are often ignored in plasmonic trapping.

Here, we experimentally quantify the influence of the thermophoretic force in plasmonic trapping and disentangle its contribution from the optical gradient force. We take advantage of the fact that the thermophilic or thermophobic behavior of a nano-object (governing the influence of the thermophoretic force) is not an intrinsic characteristic of the nano-object, but it depends on external parameters, notably the choice of the surfactant or the ionic strength.[50,56,62–64] We use different surfactant conditions to tune the behavior of 28 nm polystyrene nanoparticles between thermophobic and thermophilic. Our experimental data reveal that the surfactant plays a major role in the trap performance: a 20 × larger stiffness is measured with sodium dodecyl sulfate (SDS) as compared to Triton X-100. We relate this difference to the fact that with SDS the nanoparticles become thermophilic, and as they move towards the plasmonic hot spot, the thermophoretic force contributes constructively to the net restoring force. On the contrary with Triton, the nanoparticles are thermophobic: the thermophoretic force pushes the nanoparticles away from the trap, which lowers the trap potential. Revealing here the hidden role of the surfactant provides new directions to better understand and optimize plasmonic trapping.

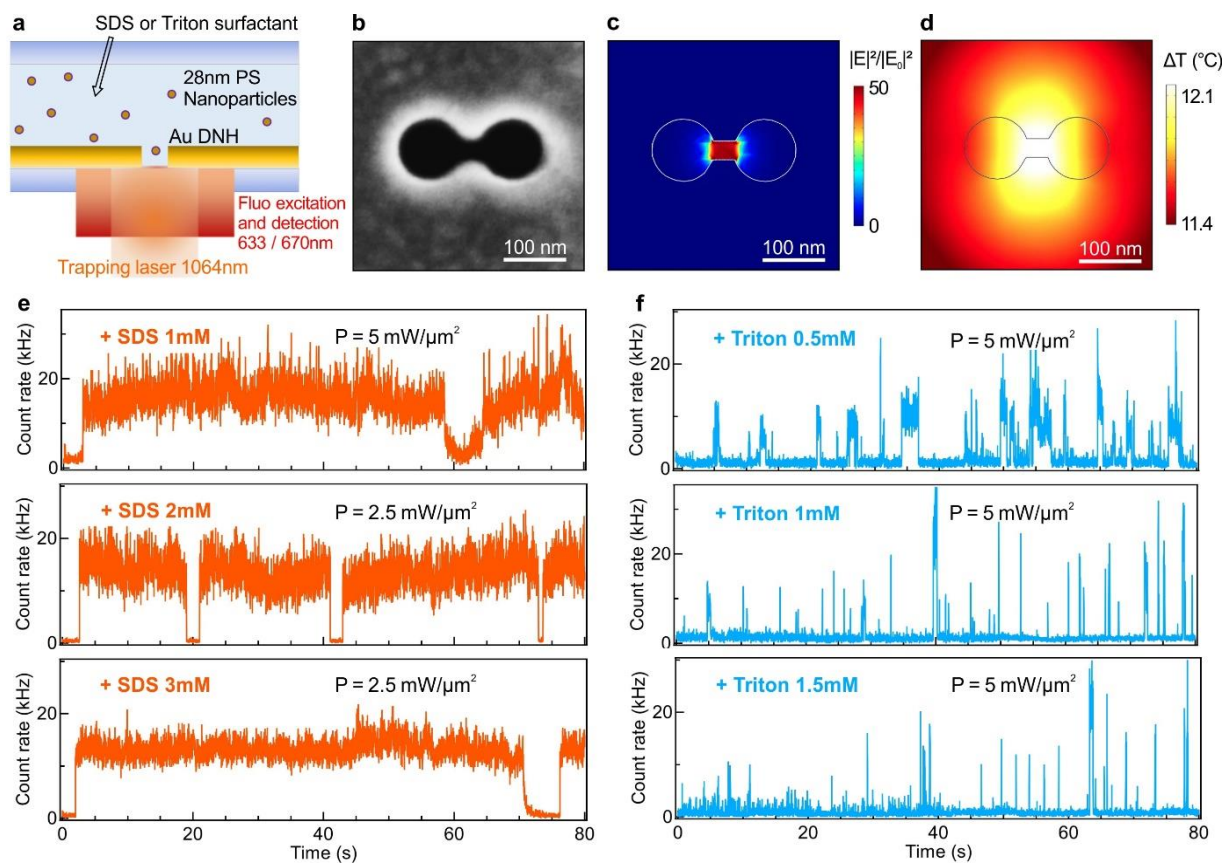

**Figure 1.** Investigating the influence of the surfactant in plasmonic nano-optical trapping. (a) Scheme of the optical setup combining an infrared laser beam for trapping with a fluorescence microscope. (b) Scanning electron microscopy image of a double nanohole aperture with 100 nm hole diameter, 30 nm gap width and 30 nm gap length. (c)-(d) Numerical simulations of the intensity enhancement in the middle of the gold layer and the temperature increase for 5 mW/µm² illumination at 1064 nm with linear vertical polarization. (e,f) Fluorescence time traces for different surfactant concentrations and infrared laser powers.

Figure 1a shows a sketch of our experiment using a double-nanohole (DNH) aperture illuminated by a 1064 nm infrared (IR) laser beam focused on a 1 µm² spot. As trapping object, we select red-fluorescent carboxylate-modified polystyrene nanoparticles of 28 nm diameter (Thermofisher reference F8783). The nanoparticles are diluted in ultrapure water with different surfactants, and no salts are added to the solution, as commonly done in plasmonic trapping to avoid nanoparticle aggregation.[2,3,7,8,12,14,17,20] Instead of using the classical transmission of the infrared laser to monitor trapping,[2–4] we take advantage of the fluorescence emission of the nanoparticles which enables to work on a nearly-zero background,[65,66] as opposed to transmission measurements which rely on tiny changes of the transmission by a few percent. Importantly in our experiments, the height of the liquid solution containing the nanoparticles on top of the DNH is set to be around 20 µm. This low height ensures that thermally-induced convection does not play a role in our observations,[45,63,64,67] contrarily to some other configurations where convection flows are used to improve the loading capacity of the trap.[61,68]

A typical DNH structure used in this work is shown in Fig. 1b. Two holes of 100 nm diameter are connected with a gap of 30 nm width and 30 nm length. The DNH is milled by focused ion beam in a 100 nm thick gold film with a 5 nm chromium adhesion layer. This geometry is optimized to yield the best performance around 1064 nm,[69] with 50× local intensity enhancement when the infrared laser polarization is set parallel to the metal apex between the holes (Fig. 1c). The temperature distribution around the DNH is computed numerically (Fig. 1d). In addition to these simulations, we take advantage of the temperature measurements and the extensive characterization of single and double nanohole structures that we have performed recently.[46,47,70] At 5 mW/µm² infrared illumination, we measure a local temperature increase of +12 °C in the DNH and a linear dependence of the temperature increase on the infrared laser power (Supporting Information Fig. S1).

For the trapping experiments, the only varying parameters are the nature and the concentration of the surfactant added to the solution, and the trapping laser power. All the other experimental conditions remain identical. As surfactants, we choose sodium dodecyl sulfate (SDS) and Triton X-100 (Triton).[56,63,64] For the concentrations used here, we check that the surfactant itself does not induce artifacts in our measurements: at 3 mM SDS or 1.5 mM Triton, similar nanoparticle concentrations and diffusion times are measured by fluorescence correlation spectroscopy (FCS) as in the absence of surfactant (Fig. S2). This ensures that the nanoparticles stay well mono-dispersed and that the presence of SDS or Triton does not induce aggregation of the nanoparticles. SDS is anionic while Triton is nonionic, therefore we exclude any electrostatic interaction with our negatively-charged carboxylate-modified polystyrene nanoparticles. We also check that the presence of the 1064 nm laser does not affect the nanoparticle fluorescence emission (Fig. S2).

Figures 1e,f directly compare trapping experiments performed in the presence of SDS or Triton. With SDS, clear jumps of the fluorescence signal are observed, indicating that a nanoparticle is being trapped. The trapping events with SDS last several tens of seconds before the nanoparticle eventually escapes the trap. Higher SDS concentrations lead to longer trapping times (Fig. 1e). The data recorded with Triton is strikingly different: only short lived bursts, typically lasting about 1 second, are observed, showing that the presence of Triton disturbs the trap potential. These events tend to disappear when the Triton concentration is increased (Fig. 1f, S3-S5). We have also checked that the fluorescence signal vanishes when the trapping laser is blocked, showing that the nanoparticle is not electrostatically adsorbed on the DNH surface, see Fig. S6.

We measure the trap stiffness by analyzing the temporal correlation of the fluorescence intensity recorded during a trapping event.[7,71–73] The theory behind this approach is thoroughly described in the Supporting Information Section S4. Briefly, for a trapped fluorescent nanoparticle, we show that the temporal correlation of the fluorescence intensity decays as $\exp(-2t/\tau)$. The correlation time $\tau$ is given by the ratio of the Stokes drag coefficient $\gamma$ and the trap stiffness $\kappa$:

$$\tau = \frac{\gamma}{\kappa} \qquad (1)$$

The presence of the DNH walls modify the Stokes drag coefficient, this effect is taken into account by applying Faxen's law:[7,74]

$$\gamma = \frac{6\pi\eta R}{\left(1 - \frac{9}{16}\frac{R}{h} + \frac{1}{8}\left(\frac{R}{h}\right)^3 - \frac{45}{256}\left(\frac{R}{h}\right)^4 - \frac{1}{16}\left(\frac{R}{h}\right)^5\right)} \qquad (2)$$

where $h$ is the average distance between the center of the nanoparticle and the aperture wall, $R$ is the nanoparticle radius and $\eta$ the viscosity of the medium. By assuming that $h$ amounts to 15 nm, the $6\pi\eta R$ term in the drag coefficient is increased by a factor 2.5 (factors between 1.6 and 3 were used in the previous work by Kotnala and Gordon,[7] our value stands well within this range). The temperature dependence of the water viscosity $\eta$ is also taken into account (Fig. S1).

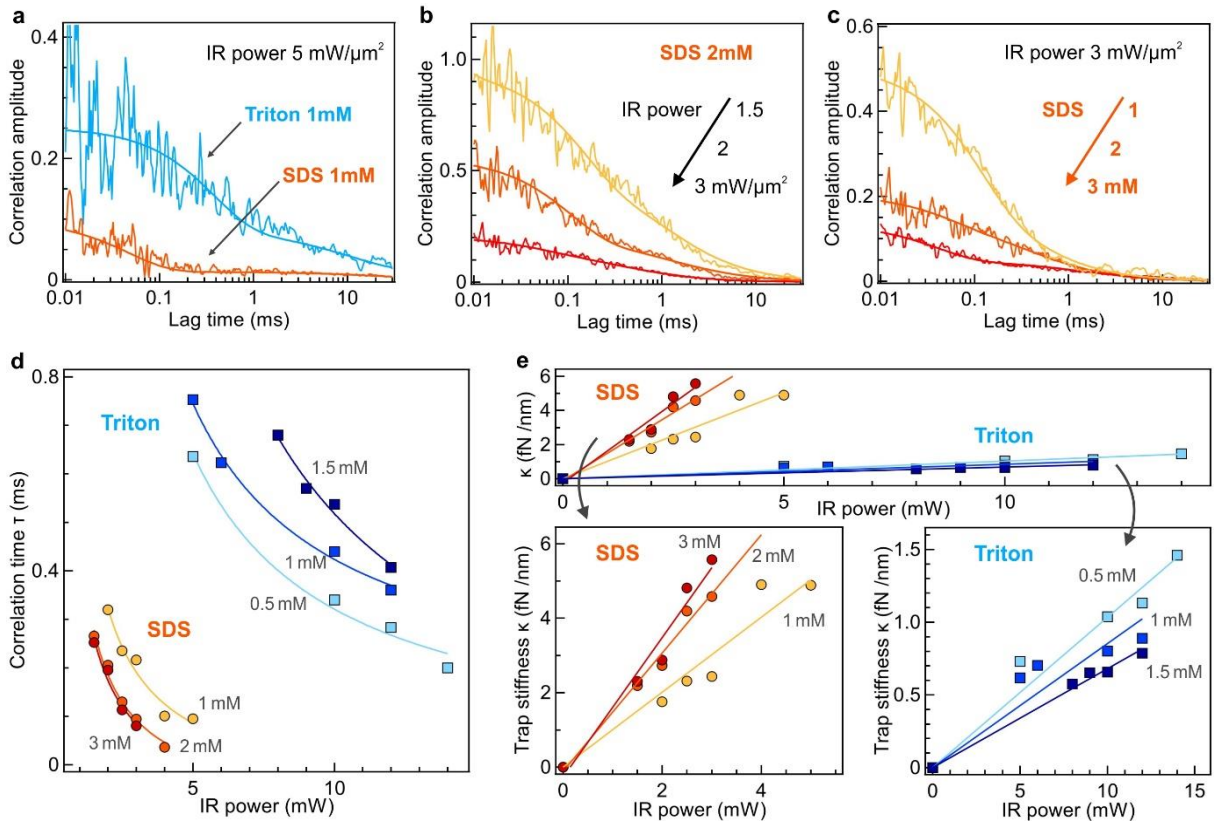

**Figure 2.** Quantifying the trap stiffness. (a) Correlation functions (thin lines) and numerical fits (thick lines) computed from the fluorescence intensity recorded during a trapping event in presence of 1 mM SDS or 1 mM Triton. (b,c) Evolution of the fluorescence correlation for different trap powers (b) and different SDS concentrations (c). (d) Correlation times extracted from the numerical fits in (a-c) as a function of the trap power. The trap stiffness $\kappa$ is deduced from the correlation time $\tau$ by $\tau = \gamma/\kappa$, where $\gamma$ is the Stokes drag coefficient. (e) Evolution of the trap stiffness as a function of the trap power. Lines are numerical fits.

Figures 2a-c show the experimental correlation data recorded for a single polystyrene nanoparticle trapped in a DNH with different surfactant conditions and different trapping powers. A higher correlation amplitude and a longer correlation time indicate larger fluctuations of the position of the fluorescent nanoparticle inside the trap and a lower trap stiffness. On the contrary, low correlation amplitudes and short correlation times indicate that the nanoparticle goes back quickly to its equilibrium position. This corresponds to a deep and narrow trap potential. Comparing SDS and Triton (Fig. 2a), a clearly reduced correlation amplitude and faster correlation time are observed for SDS, indicating a larger trap stiffness for SDS. For a given surfactant concentration, increasing the trap power lowers the correlation amplitude and reduces the correlation time (Fig. 2b and S7a). This confirms that the higher trap stiffness (due to the higher trap power) indeed leads to lower correlation amplitudes and correlation times. We also change the surfactant concentration while the infrared

power is fixed. Increasing the SDS concentration reduces both the correlation amplitude and characteristic time (Fig. 2c), validating the positive effect of SDS on the trap potential. We confirm Triton's negative role as higher Triton concentrations increase the correlation amplitude and characteristic time (Fig. S7b).

Numerical interpolation of the intensity correlation data quantifies the correlation time $\tau$, the results are summarized in Fig. 2d. Then the trap stiffness is computed as $\kappa = \gamma/\tau$. For all SDS and Triton concentrations, we find that the trap stiffness follows the expected linear dependence with the trap power (Fig. 2e). From the slope of this line, we determine the intensity-normalized trap stiffness in fN/nm/mW. For 1 mM SDS, the DNH yields a stiffness of 1.0 ± 0.1 fN/nm/mW while for 0.5 mM Triton, the stiffness is about 10× lower at 0.10 ± 0.01 fN/nm/mW. This large difference highlights the importance of the surfactant in the plasmonic trap performance. Comparing our values with works from other groups, a stiffness of 0.1 fN/nm/mW was reported for 20 nm polystyrene nanoparticles trapped in a DNH.[7] The nanoparticles are suspended in water (no surfactant case, which appears similar to 0.5mM Triton in our experiments). Simulations for coaxial apertures indicate trap stiffness of 0.36 fN/nm/mW for 20 nm polystyrene nanoparticles,[19] while connected nanohole arrays achieved 0.85 fN/nm/mW for 30 nm polystyrene nanoparticles in presence of 0.1% Tween 20 surfactant.[20] These different results confirm the consistency of our values for the trap stiffness. Altogether, the results in Fig. 1 and 2 show that the surfactant is a powerful understated way to improve the nano-optical trap performance without changing much the experiment design nor the plasmonic nanostructure itself. We control that the nature of the gold surface does not influence our observations: covering the gold surface with a self-assembled dithiothreitol monolayer yields similar results to an untreated sample (Fig. S8).

To go one step further, we need to characterize the thermophilic or thermophibic response of the nanoparticles. For this purpose, we switch to an array of gold nanoparticles uniformly deposited on a glass substrate using block copolymer micellar lithography (Fig. 3a, Fig. S9b).[75,76] This sample enables direct observation of the fluorescent signal from the polystyrene nanoparticles. We carefully set the infrared illumination condition to achieve a similar temperature increase as with the DNH experiment (Fig. S9).[46,75] For the gold nanoparticle array, the collimated infrared illumination intensity (Fig. S9b) is 1000× less than for the DNH trapping, so no gradient optical force is present in these experiments. A sCMOS camera records the fluorescence image of the sample which directly reflects the local concentration of the fluorescent nanoparticles. This scheme allows to follow the thermophoretic movement of nanoparticles upon infrared illumination heating.

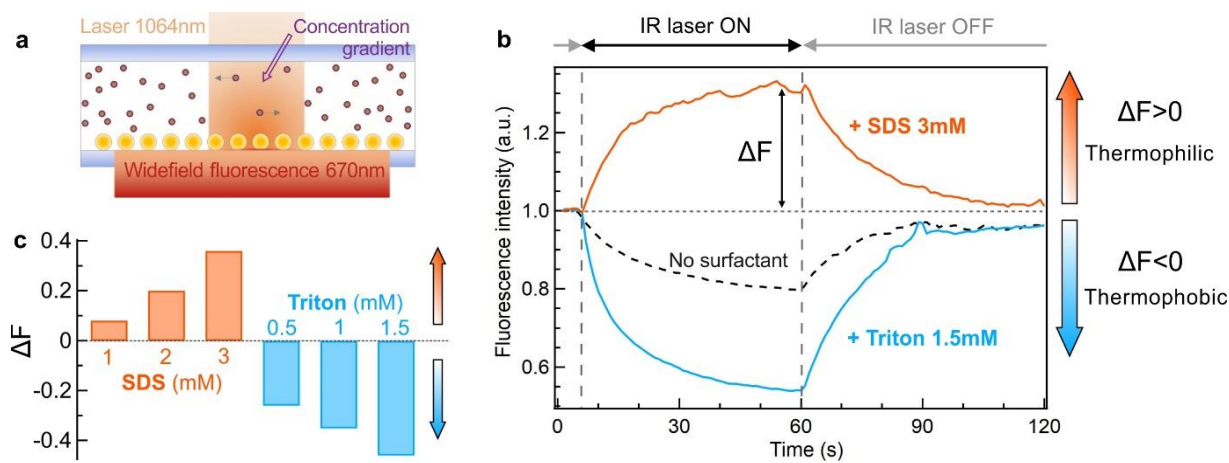

**Figure 3.** The thermophilic/thermophobic response of the nanoparticles is controlled by the surfactant. (a) Sketch of the experiment for fluorescence imaging. (b) Temporal evolution of the normalized fluorescence intensity when the heating infrared laser is switched on and off. An intensity increase correspond to a thermophilic behavior as more nanoparticles gather towards the infrared laser spot. Conversely, a fluorescence decrease show a thermophobic response. (c) The maximum relative gain or loss ΔF in the fluorescence intensity is controlled by the surfactant.

Figure 3b shows the evolution of the fluorescence intensity in the center of the spot when the infrared laser is turned on (at t = 6 s) and off (at t = 60 s). While all other experimental parameters are strictly identical, in the presence of SDS, we monitor an increase of the fluorescence intensity upon heating, while for Triton the fluorescence decreases (additional traces are shown in Fig. S9). As important check for the dynamic nature of the process (no static agglomeration of nanoparticles), the fluorescence signal goes back to its initial level when the infrared laser is turned off. The fluorescence gain with SDS is related to an increase of the local concentration of fluorescent nanoparticles which tend to gather towards the hot region, featuring a thermophilic behavior. Conversely with Triton, the fluorescence loss corresponds to a drop of the number of nanoparticles in the hot spot as the nanoparticles move away from the hot region with a thermophobic behavior. These experimental results support the earlier reports indicating that 22 nm polystyrene nanoparticles are thermophobic at room temperature in presence of 1 mM Triton,[63] while SDS can lead to a thermophilic response.[62,64] The thermal transport of charged colloids depends on the nature of the particle-solvent interface. We refer the reader to a recent review for a detailed discussion of the origin of this effect.[77]

To quantify the gain or loss of the fluorescence signal, we introduce the quantity ΔF as defined on Fig. 3b. This relative fluorescence intensity change directly reflects the thermophilic/thermophobic response of the nanoparticles in presence of the surfactant. A positive ΔF denotes thermophilic (particles moves towards the hot), while a negative ΔF corresponds to thermophobic. The evolution of ΔF nicely depends linearly on the surfactant concentration, allowing to easily tune the thermophilic/thermophobic response of the nanoparticles (Fig. 3c).

Now we can gather together the intensity-normalized trap stiffness $\kappa$ with the thermophilic/thermophobic response ΔF. Figure 4a shows a scatter plot of these two quantities, indicating a clear correlation between the thermophobic/thermophilic response and the trap stiffness. We discuss in the Supporting Information section S8 the link between the thermophoretic stiffness and the fluorescence change and how these quantities both relate to the Soret coefficient. Having a quantitative measurement for the total trap stiffness $\kappa$, we can decompose it into the contribution $\kappa_{opt}$ stemming from the optical gradient force and the part $\kappa_{therm}$ from the thermophoretic force (Fig. 4b). The contribution from the optical gradient force is assumed to be independent of the surfactant presence (the Supporting Information Fig. S2 checks that the surfactant concentrations used here have a minimal influence on the medium viscosity). Hence we estimate $\kappa_{opt}$ from the interpolation of the data in Fig. 4a when ΔF =0 and there is no thermophoretic influence (Fig. S10). This approach provides a value of $\kappa_{opt}$ = 0.6 ± 0.1 fN/nm/mW, which appears well in line with recent experimental and theoretical works.[7,12,19,20,52] Then knowing the total stiffness $\kappa$ and the optical contribution $\kappa_{opt}$, we deduce the thermophoretic contribution as $\kappa_{therm} = \kappa - \kappa_{opt}$ (Fig. 4b). This thermophoretic stiffness depends clearly on the surfactant, featuring positive values for SDS and negative values for Triton. For SDS (thermophilic nanoparticles), the thermophoretic force will positively add to the optical force and increase the overall trap stiffness. On the contrary for Triton (thermophobic nanoparticles), the thermophoretic force has an opposite sign to the optical force and will counterbalance the optical gradient influence, thereby lowering the overall trap stiffness. It is important to notice that the stiffness contribution from the thermophoretic force $\kappa_{therm}$ has an amplitude that is similar to the optical contribution $\kappa_{opt}$: we find $\kappa_{therm}$ = 0.4 ± 0.1 fN/nm/mW for 1mM SDS and $\kappa_{therm}$ = -0.5 ± 0.1 fN/nm/mW for 1mM Triton.

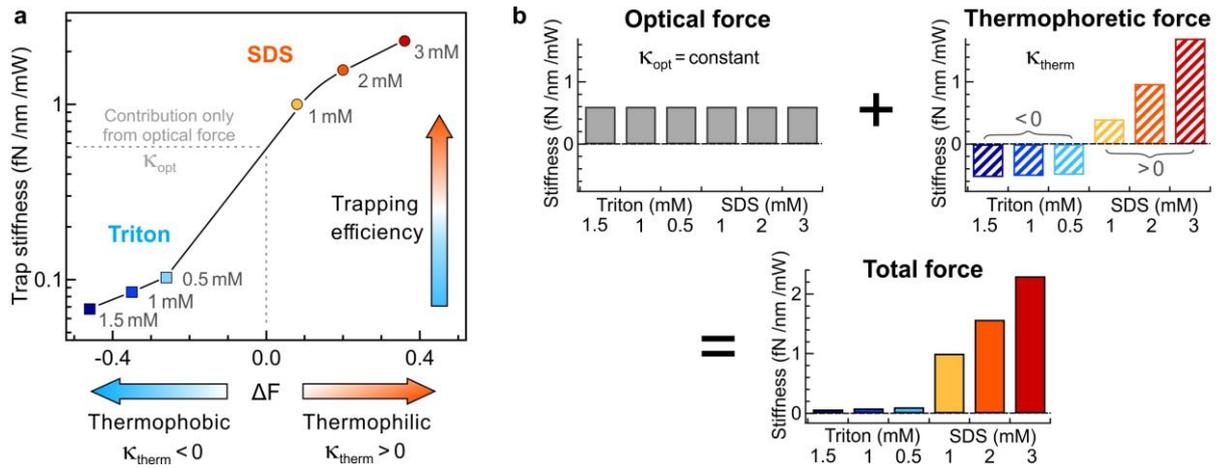

**Figure 4.** Quantifying the surfactant role and the thermophoretic force. (a) Plot of the intensity-normalized trap stiffness as a function of ΔF. The line is a guide to the eyes. From the value extrapolated for ΔF =0 (no thermophoretic force), we deduce a trap stiffness $\kappa_{opt}$ which we attribute only to the optical force. (b) Decomposition of the total trap stiffness as a sum of the stiffness due to the optical force $\kappa_{opt}$ (independent on the surfactant) and the one attributed to the thermophoretic force $\kappa_{therm}$ (surfactant-dependent, positive for thermophilic behavior, negative for thermophobic).

In conclusion, we have demonstrated the unrecognized influence played by the surfactant and the thermophoretic force in plasmonic nano-optical trapping. While all the other experimental conditions remain exactly identical, the nature and the concentration of the surfactant can have a dramatic effect on the trap performance: up to 20 × larger trap stiffness is obtained with 3 mM SDS as compared to 1.5 mM Triton. In presence of SDS, the nanoparticles are thermophilic, and the thermophoretic force adds positively to the optical gradient force, increasing the overall trap stiffness. For Triton however, the nanoparticles become thermophobic, leading to a thermophoretic force with opposite sign to the optical gradient force, lowering the net trap potential. This set of experiments disentangles the thermophoretic contribution from the optical gradient force in plasmonic nano-tweezers. Our procedure can be easily extended to investigate other plasmonic geometries. Altogether, our results help to better understand nano-optical trapping experiments and improve the trap performance by optimizing the surfactant conditions.

**ASSOCIATED CONTENT**

**Supporting Information**

Temperature increase in the DNH and water viscosity calibration, Control that the surfactant does not affect the nanoparticle diffusion or concentration, Supplementary fluorescence time traces with increasing infrared powers, Trap stiffness quantification from the fluorescence intensity time trace, Additional correlation data with Triton, Correlation analysis in presence of a self-assembled monolayer on the gold surface, Temporal evolution of the fluorescence signal depending on the thermophilic / thermophobic response of the nanoparticles, Relationship between the thermophoretic stiffness and the fluorescence change, Linear-linear plot of the intensity-normalized trap stiffness as a function of ΔF, Additional method details.

The Supporting Information is available free of charge on the ACS Publications website at DOI: xxxxxxxxx

**Funding Sources**

This project has received funding from the European Research Council (ERC) under the European Union's Horizon 2020 research and innovation programme (grant agreements No 723241 TryptoBoost and No 772725 HiPhore) and from the Agence Nationale de la Recherche (ANR) under grant agreement ANR-17-CE09-0026-01 and ANR-18-CE42-0013.

**Conflict of Interest**

The authors declare no competing financial interest.

# Supporting Information for

# Quantifying the Role of the Surfactant and the Thermophoretic Force in Plasmonic Nano-Optical Trapping

Quanbo Jiang,[1] Benoît Rogez,[1] Jean-Benoît Claude,[1] Guillaume Baffou,[1] Jérôme Wenger[1,*]

[1] Aix Marseille Univ, CNRS, Centrale Marseille, Institut Fresnel, 13013 Marseille, France

* Corresponding author: jerome.wenger@fresnel.fr

This document contains the following supporting information:

S1. Temperature increase in the DNH and water viscosity calibration

S2. Control that the surfactant does not affect the nanoparticle diffusion or concentration

S3. Supplementary fluorescence time traces

S4. Trap stiffness quantification from the fluorescence intensity time trace

S5. Additional correlation data with Triton

S6. Correlation analysis in presence of a self-assembled monolayer on the gold surface

S7. Temporal evolution of the fluorescence signal depending on the thermophilic / thermophobic response of the nanoparticles

S8. Relationship between the thermophoretic stiffness and the fluorescence change

S9. Linear-linear plot of the intensity-normalized trap stiffness as a function of ΔF

S10. Additional method details

**S1. Temperature increase in the DNH and water viscosity calibration**

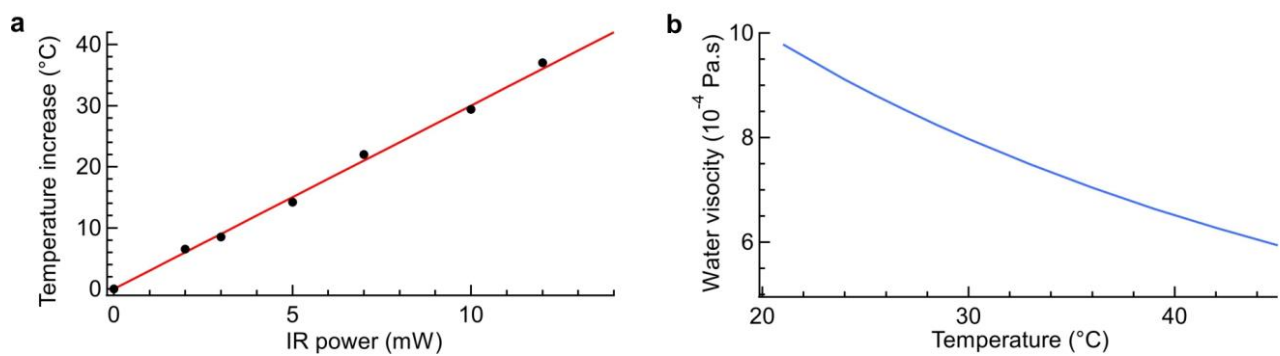

**Figure S1.** (a) Temperature increase in the DNH as a function of the infrared laser power. The temperature measurements use the same lifetime-based procedure as described in our previous works.[1,2] Dots are experimental measurements, the red line is a numerical fit. Here, the adhesion layer for the gold film is 5 nm Cr which matches well the trend observed previously with different adhesion layer materials and thicknesses.[2] (b) In the calculation of the Stokes drag coefficient $\gamma$, we use the viscosity of the medium $\eta$, which we take to be the temperature-dependent viscosity of water calculated using the Vogel equation.[3] Hence, for any given infrared laser power, we know from (a) the local temperature and from (b) we deduce the corresponding viscosity for water.

## S2. Control that the surfactant does not affect the nanoparticle diffusion or concentration

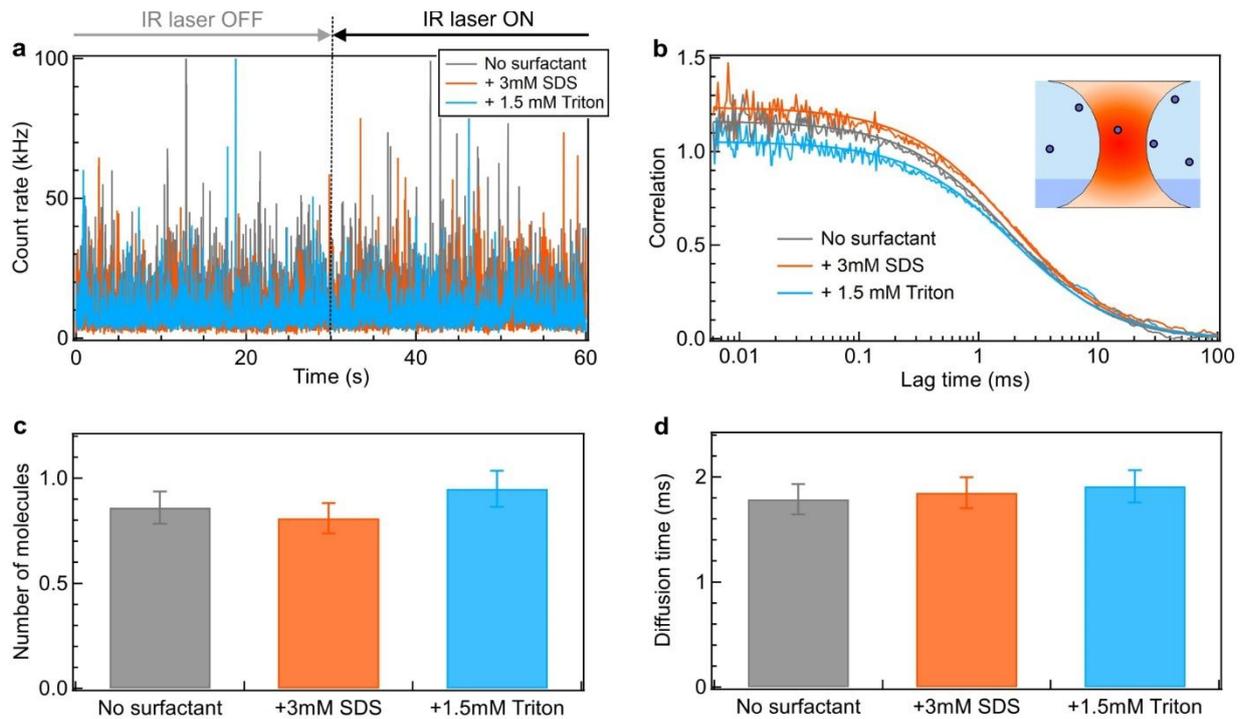

**Figure S2.** Control that the surfactant alone does not modify the Brownian diffusion properties of the nanoparticles and does not induce aggregation or agglomeration of the nanoparticles. This set of experiments is performed within a water-based solution on top of a glass coverslip using a confocal microscope. There is no gold and hence no temperature gradient and no thermophoretic force here. The surfactant concentrations correspond to the highest used in this work: 3mM for SDS and 1.5mM for TritonX-100. (a) Fluorescence time traces of the 28 nm nanoparticles. The 635 nm laser intensity used to excite the nanoparticle fluorescence is 5 µW/µm², while the 1064 nm laser intensity is set to 10 mW/µm² as for the trapping experiments. At this power level, no influence of the infrared laser beam is seen neither on the fluorescence photophysics nor on the nanoparticles diffusion properties. The fluorescence bursts stemming from the diffusion of single nanoparticles across the confocal detection volume have similar peak intensities and burst durations. (b) FCS analysis of the time traces in (a). From the numerical interpolation of this data using a standard 3D Brownian diffusion model,[4] we deduce the average number of nanoparticles in the confocal detection volume (c) and their FCS diffusion time (d). Both the number of nanoparticles and their diffusion time appear to be unaffected by the presence of the surfactant.

## S3. Supplementary fluorescence time traces

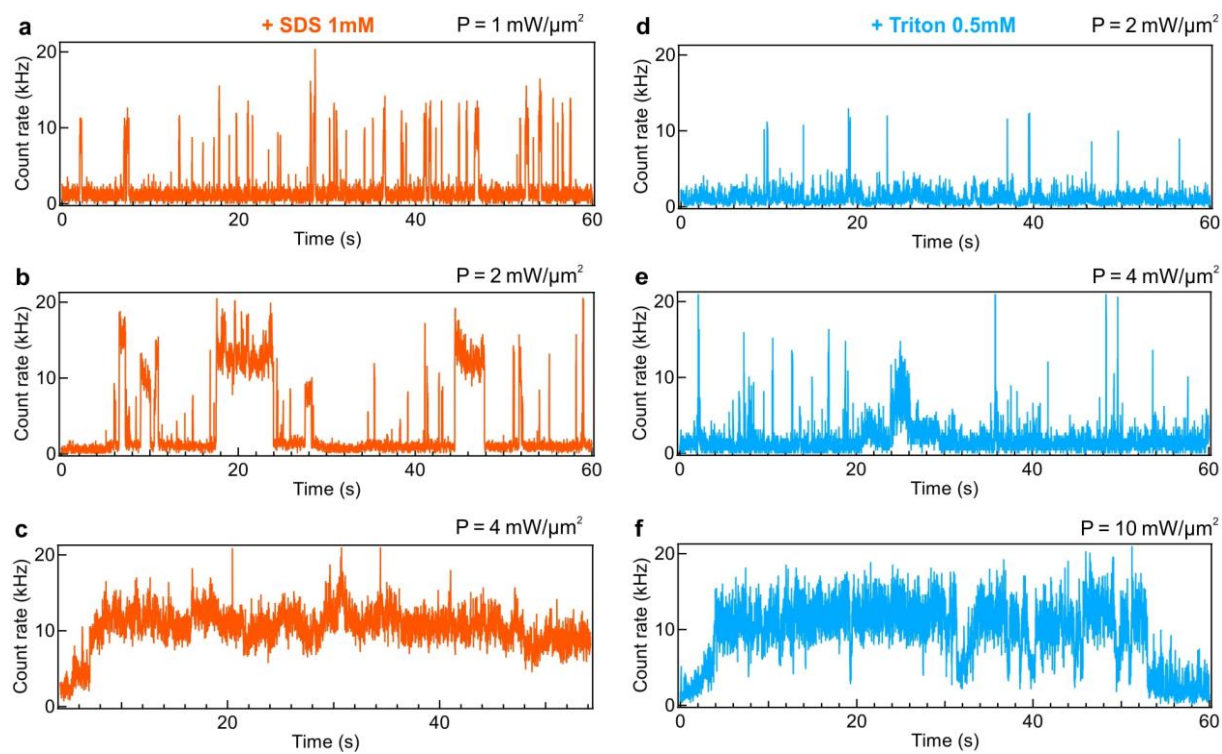

**Figure S3.** Fluorescence time traces for 1mM SDS (a-c) and 0.5 mM Triton (d-f) recorded with increasing infrared trap powers. While 2mW/µm² IR intensity is enough to start observing nanoparticle trapping events with 1mM SDS, in the presence of Triton, about 2 to 3× higher powers are required.

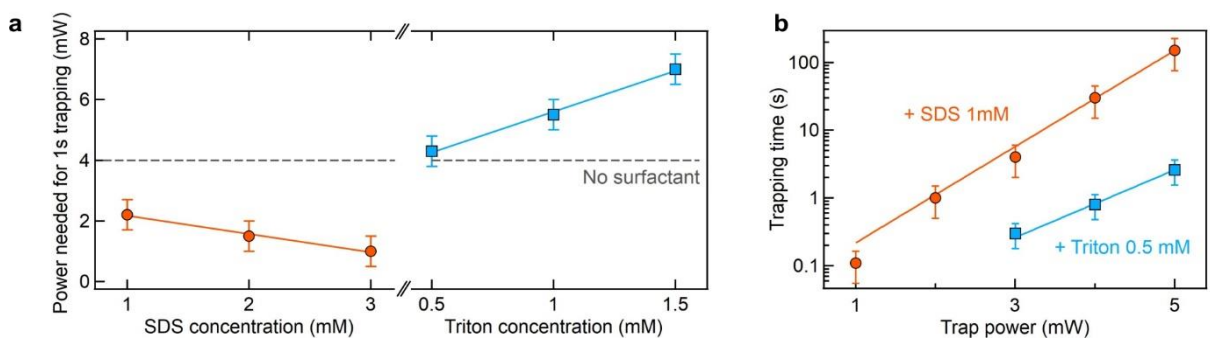

**Figure S4.** (a) Comparison of the minimum infrared laser powers needed to observe trapping events lasting more than 1 second as a function of the nature and the concentration of the surfactant. The dashed horizontal line indicates the power level when no surfactant is added to the solution. (b) Evolution of the trapping time as a function of the infrared laser power. Lines are exponential fits following $\exp(E_{trap}/k_B T)$, where $E_{trap}$ is the trap potential which is proportional to the infrared laser power.

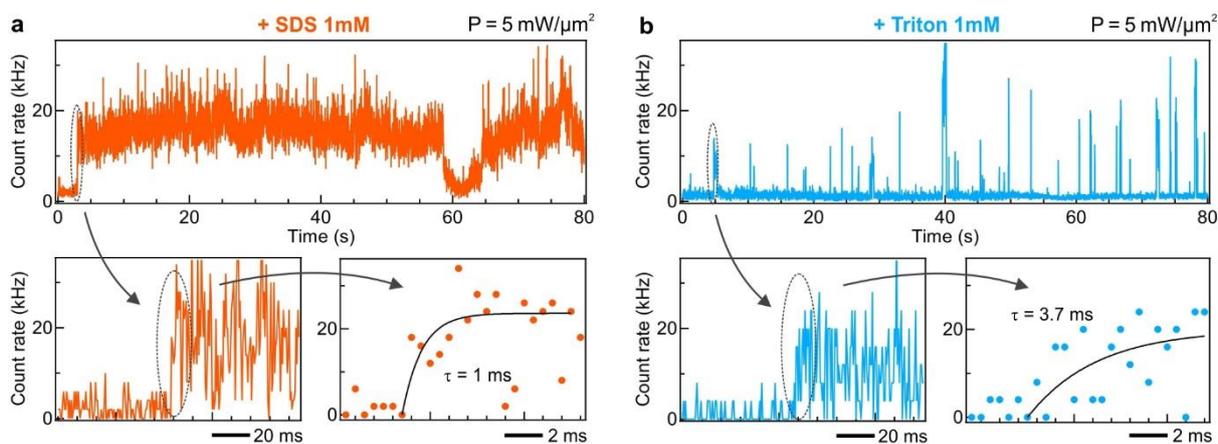

**Figure S5.** (a) Fluorescence intensity time traces recorded while trapping individual 28 nm fluorescent polystyrene nanoparticles with either 1 mM SDS (a) or Triton (b) under otherwise identical conditions. The insets show a close-up view of a trapping event together with an exponential fit. The bin time is 0.5 ms for the expanded views and 10 ms for the other traces. The transition time when the nanoparticle enters the trap appears to be significantly longer in the presence of Triton (3.7 ms) as compared to SDS (estimated to 1 ms, here we are limited by the 0.5 ms time resolution of our instrument). This longer transition time is another indication that the DNH trap stiffness differs between SDS and Triton.

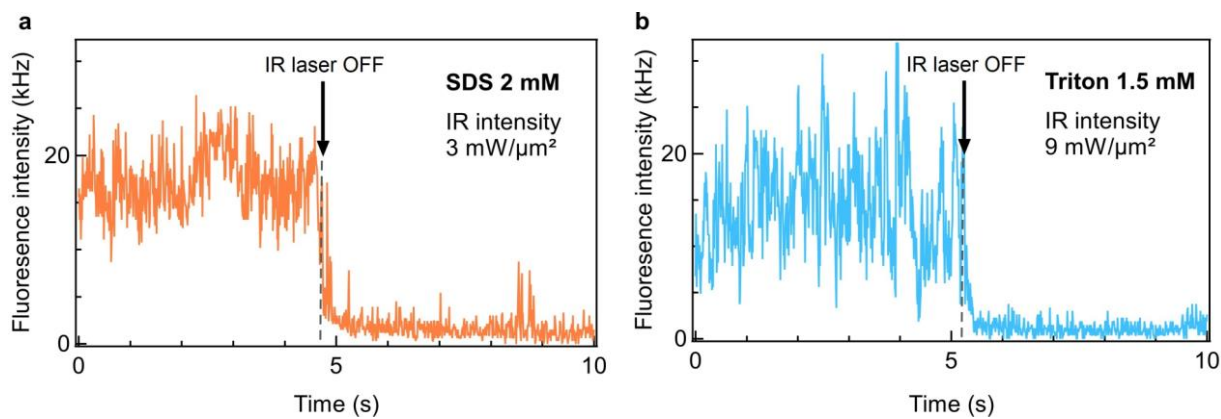

**Figure S6.** Fluorescence time traces showing that the nanoparticle is released (drop of fluorescence signal) when the infrared trapping laser is blocked for 2 mM SDS (a) and 1.5 mM Triton (b). The red laser for fluorescence excitation is constantly on during the whole experiment.

**S4. Trap stiffness quantification from the fluorescence intensity time trace**

Here we derive the theoretical framework used to determine the trap stiffness from the fluorescence signal time traces.[5–8] The core idea of the method is to analyze the temporal fluctuations of the fluorescence intensity $I(t) = I(r(t))$ which varies in time as the trapped nanoparticle explores different positions $r(t)$ inside the trap. For simplicity, we assume that the trapped nanoparticle is a point-like source. Its fluorescence emission inside the trap follows a spatial distribution $I(r)$ centered around the position $r = 0$. The intensity distribution can be expanded along the three directions $i = x, y, z$ as:[7]

$$I(r(t)) = I_0 + \frac{1}{2}\sum_i (\frac{\partial^2 I}{\partial r_i^2}) r_i(t)^2 = I_0(1 - \sum_i \frac{r_i(t)^2}{2\omega_i^2}) \quad (S1)$$

where $r(t)$ is the bead displacement from the equilibrium center position (maximum fluorescence) which is actually time-dependent, $I_0$ is the peak intensity and $\omega_i$ is the width at 1/e² of the intensity distribution around the peak in the center of the DNH nanogap.

To analyze the fluctuations of the fluorescence intensity, we compute its temporal correlation function $\langle I(0)I(t) \rangle$, where $\langle \rangle$ denotes time averaging. Using the expansion in Eq. (S1), the temporal correlation is given by:[6,7]

$$\langle I(r(0))I(r(t)) \rangle \cong I_0^2 \left[ (1 - \frac{1}{2}\sum_i \frac{\langle r_i^2(t) \rangle}{\omega_i^2})^2 + \frac{1}{4}\sum_i \frac{1}{\omega_i^4}(\langle r_i^2(t) r_i^2(0) \rangle - \langle r_i^2(t) \rangle^2) \right] \quad (S2)$$

We assume that inside the trap the nanoparticle position follows a Gaussian statistics, hence the last term in the equation above can be simplified as $\langle r_i^2(t) r_i^2(0) \rangle - \langle r_i^2(t) \rangle^2 = 2\langle r_i(0) r_i(t) \rangle^2$:

$$\langle I(0)I(t) \rangle \cong I_0^2 \left[ (1 - \frac{1}{2}\sum_i \frac{\langle r_i^2 \rangle}{\omega_i^2})^2 + \frac{1}{2}\sum_i \frac{1}{\omega_i^4}\langle r_i(0) r_i(t) \rangle^2 \right] \quad (S3)$$

As $\langle r_i^2(t) \rangle = \langle r_i^2 \rangle$ is time independent, the fluorescence intensity correlation $\langle I(0)I(t) \rangle$ is directly related to the nanoparticle position correlation $\langle r_i(0) r_i(t) \rangle$.[6,7]

To calculate the position correlation $\langle r_i(0) r_i(t) \rangle$, we assume that the nanoparticle undergoes Brownian motion inside a harmonic potential $V$. The position of the nanoparticle obeys the Langevin equation:

$$m \frac{d^2 r_i}{dt^2} = -\gamma \frac{dr_i}{dt} - \frac{\partial V}{\partial r_i} + f(t) \quad (S4)$$

where $\gamma$ is the friction constant or Stokes drag coefficient and $f(t)$ represents the sum of the forces due to the incessant collision of the fluid molecules with the Brownian particles which is treated as an stochastic uncorrelated thermal force.[9] $f(t)$ follows a Gaussian probability distribution with $\langle f(t) \rangle = 0$ and $\langle f(t_1) f(t_2) \rangle = 2\gamma k_B T \delta(t_1 - t_2)$. We define the trap potential as $V = \frac{1}{2}\kappa_i r_i^2$ where $\kappa_i$ represents the force constant or effective trap stiffness along each direction. Neglecting the inertial term in Eq. (S4), the Langevin equation can be rewritten as

$$\gamma \frac{dr_i}{dt} + \kappa_i r_i = f(t) \quad (S5)$$

This equation can be solved to express $r_i(t)$ in terms of $f(t')$:[10]

$$r_i(t) = \frac{1}{\gamma} \int_{-\infty}^{t} dt' \exp\left(-\frac{t-t'}{\tau_i}\right) f(t') \tag{S6}$$

where $\tau_i = \gamma/\kappa_i$. Then the position correlation can be computed [6–8]

$$\langle r_i(0) r_i(t) \rangle = \langle r_i^2 \rangle \exp\left(-\frac{t}{\tau_i}\right) \tag{S7}$$

and the mean square of the position amplitude is given by $\langle r_i^2 \rangle = k_B T / \kappa_i$. Inserting the position correlation Eq. (S7) into eq. (S3) finally gives the intensity correlation:

$$\langle I(0) I(t) \rangle \cong I_0^2 \left[ (1 - \frac{1}{2} \sum_i \frac{\langle r_i^2 \rangle}{\omega_i^2})^2 + \frac{1}{2} \sum_i \frac{\langle r_i^2 \rangle^2}{\omega_i^4} \exp\left(-2\frac{t}{\tau_i}\right) \right] \tag{S8}$$

In practice, it is more convenient to work with the intensity-normalized correlation function $G(t) = \langle \delta I(0) \delta I(t) \rangle / \langle I(t) \rangle^2$, where $\delta I(t) = I(t) - \langle I \rangle$ is the intensity fluctuation around the average. Assuming that the nanoparticle is trapped strongly and remains near the trap center, the mean square displacement $\langle r_i^2 \rangle$ can be neglected as compared to the trap dimensions $\omega_i$. Equation (S8) can then be further simplified to [6–8]

$$G(t) = \frac{\langle \delta I(0) \delta I(t) \rangle}{\langle I(t) \rangle^2} \cong \frac{1}{2} \sum_i \left(\frac{\langle r_i^2 \rangle}{\omega_i^2}\right)^2 \exp\left(-2\frac{t}{\tau_i}\right) \tag{S9}$$

Equation (S9) is our main result. It shows that the temporal correlation of the fluorescence intensity recorded for a trapped nanoparticle decays as $\exp(-2\frac{t}{\tau_i})$ where the correlation time is given by

$$\tau_i = \frac{\gamma}{\kappa_i} \tag{S10}$$

The Stokes drag coefficient of the solution is primarily determined by $6\pi\eta R$ where $R$ is the 14 nm nanoparticle radius and $\eta$ the viscosity of the medium (computed along Fig. S1b). To account for the presence of the DNH walls, the drag coefficient is modified by applying Faxen's law:[11]

$$\gamma = \frac{6\pi\eta R}{(1 - \frac{9}{16}\frac{R}{h} + \frac{1}{8}\left(\frac{R}{h}\right)^3 - \frac{45}{256}\left(\frac{R}{h}\right)^4 - \frac{1}{16}\left(\frac{R}{h}\right)^5)} \tag{S11}$$

where $h$ is the average distance between the center of the nanoparticle and the aperture wall. We assume that $h$ amounts to 15 nm in the gap of the DNH with dimensions 30 nm x 30 nm, so that the additional Faxen's term in Eq. (S11) increases the $6\pi\eta R$ term by a factor 2.5.[5] From the analysis of the fluorescence intensity correlation function $G$, the correlation time $\tau_i$ is deduced. Using Eqs. (S10) and (S11), the trap stiffness is finally obtained as $\kappa_i = \gamma/\tau_i$. Theoretically, the variations of the trap potential along the three different directions $i = x, y, z$ could be assessed. However, we find experimentally that the correlation functions are quite noisy for short lag times below 10 μs which makes their fitting very inaccurate. Hence to decide to focus mainly on the term in the sub-ms range which we assign to the gradient along the $z$ direction.

The correlation amplitude evolves as $\left(\frac{\langle r_i^2 \rangle}{\omega_i^2}\right)^2 = \frac{(k_B T)^2}{\kappa_i^2 \omega_i^4}$. It is therefore proportional to $\kappa_i^{-2}$ so that a higher trap stiffness (due to the surfactant or the trap power) will lead to a reduced correlation amplitude.

We find that the experimental correlation functions are quite noisy for very short lag times below 10 µs. This noise currently prevents any reliable investigation for correlation times below 10 µs as the numerical fits are very inaccurate in this range. To compare the trap stiffness for different surfactant conditions, we decide to focus on the correlation term in the sub-ms range which we assign to the gradient along the $z$ direction.

Moreover, we find that additional correlations occurring on times larger than 10 ms tend to randomly affect the data. These fluctuations may be induced by the presence of another nanoparticle around the trap which competes with the nanoparticle being trapped.[12,13] Other sources of noise such as mechanical drifts or fluorescence photobleaching may also affect the long-term stability of the fluorescence signal. The amplitude of this long-term contribution depends a lot on the selection of the time interval used to compute the correlation function. Hence we do not use it to determine the trap stiffness, but we still account for its contribution for a better fitting of the correlation data.

With these considerations, the experimental fluorescence correlation is fitted using the model:

$$G(t) = \rho_1 \exp\left(-2\frac{t}{\tau}\right) + \frac{\rho_2}{1 + \frac{t}{\tau_L}} \quad (S12)$$

where $\rho_1$ and $\rho_2$ are the correlation amplitudes, $\tau = \gamma/\kappa$ is the correlation time used to determine the trap stiffness, and $\tau_L$ is an additional correlation time used to account for the long-term time fluctuations. The first term in the right side of Eq. (S12) accounts for the nano-optical trapping, while the second term represents the additional noise fluctuations. The shape of this contribution as $1/(1 + \frac{t}{\tau_L})$ is empirically found to provide a better numerical fit than an exponential function, and this function appears reminiscent of fluorescence correlation spectroscopy in 3D systems where the detection volume along one axis is very long as compared to the two other directions.

**S5. Additional correlation data with Triton**

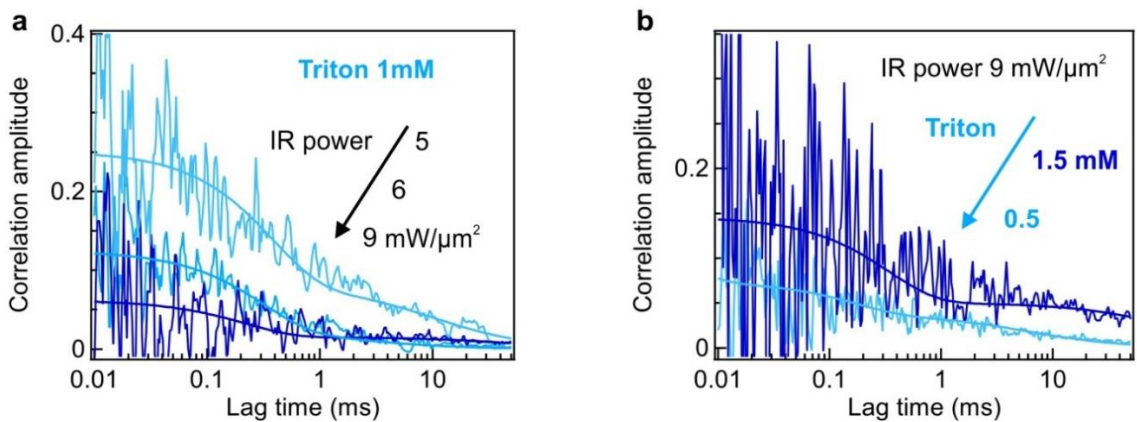

**Figure S7.** (a) Correlation functions (thin lines) and numerical fits (thick lines) computed from the fluorescence intensity recorded during a trapping event in presence of 1 mM Triton for different trap powers. (b) Influence of the Triton concentration on the correlation data for a fixed trap power.

## S6. Correlation analysis in presence of a self-assembled monolayer on the gold surface

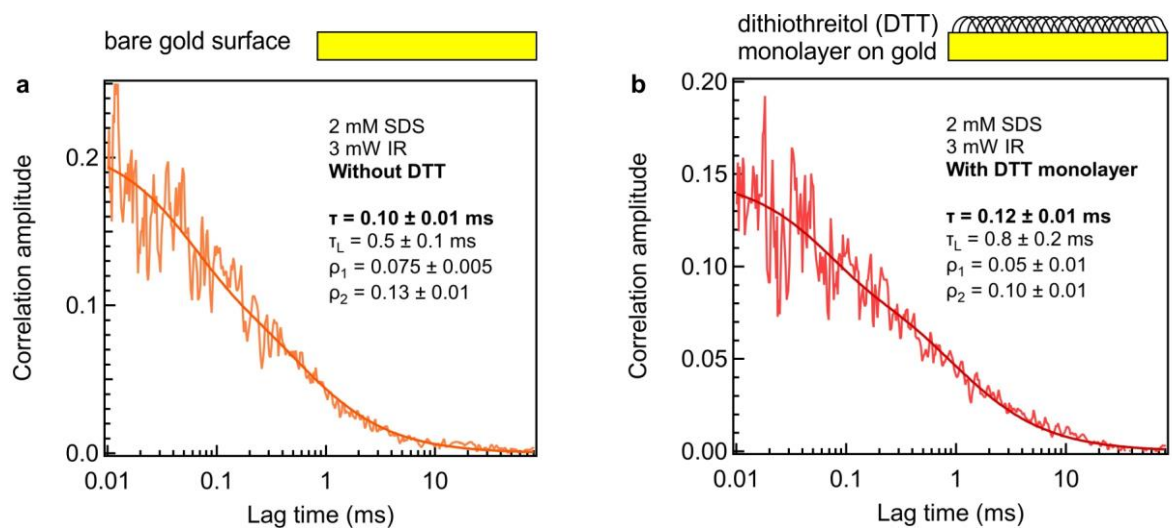

**Figure S8.** (a) Correlation functions (thin line) and numerical fits (thick line) measured for the bare gold surface (to serve as reference as in the rest of our study) in presence of 2 mM SDS at 3 mW/µm² trap power. (b) Same experimental configuration as (a) except that the gold surface has been covered by a self-assembled monolayer of dithiothreitol (DTT) before the trapping experiment.[14] Within the experimental uncertainties, both cases (a) and (b) yield similar correlation times $\tau$ and trap stiffness. Hence we can conclude that our observations are not affected by the nature of the gold surface. The presence of the self-assembled monolayer may also reduce the net concentration of SDS in the solution as some SDS molecules may bind to the self-assembled monolayer, as shown previously for a cysteamine monolayer on gold.[15]

For the DTT self-assembled monolayer formation, the gold surface is cleaned with air plasma for 20 minutes (Diener Zepto 50 W) then immediately afterwards the sample is dip into a 8 mM solution of DTT in ethanol. After 12 hours incubation at room temperature, the sample is thoroughly rinsed with ethanol to keep only a self-assembled monolayer on the gold surface.[14]

## S7. Temporal evolution of the fluorescence signal depending on the thermophilic/thermophobic response of the nanoparticles

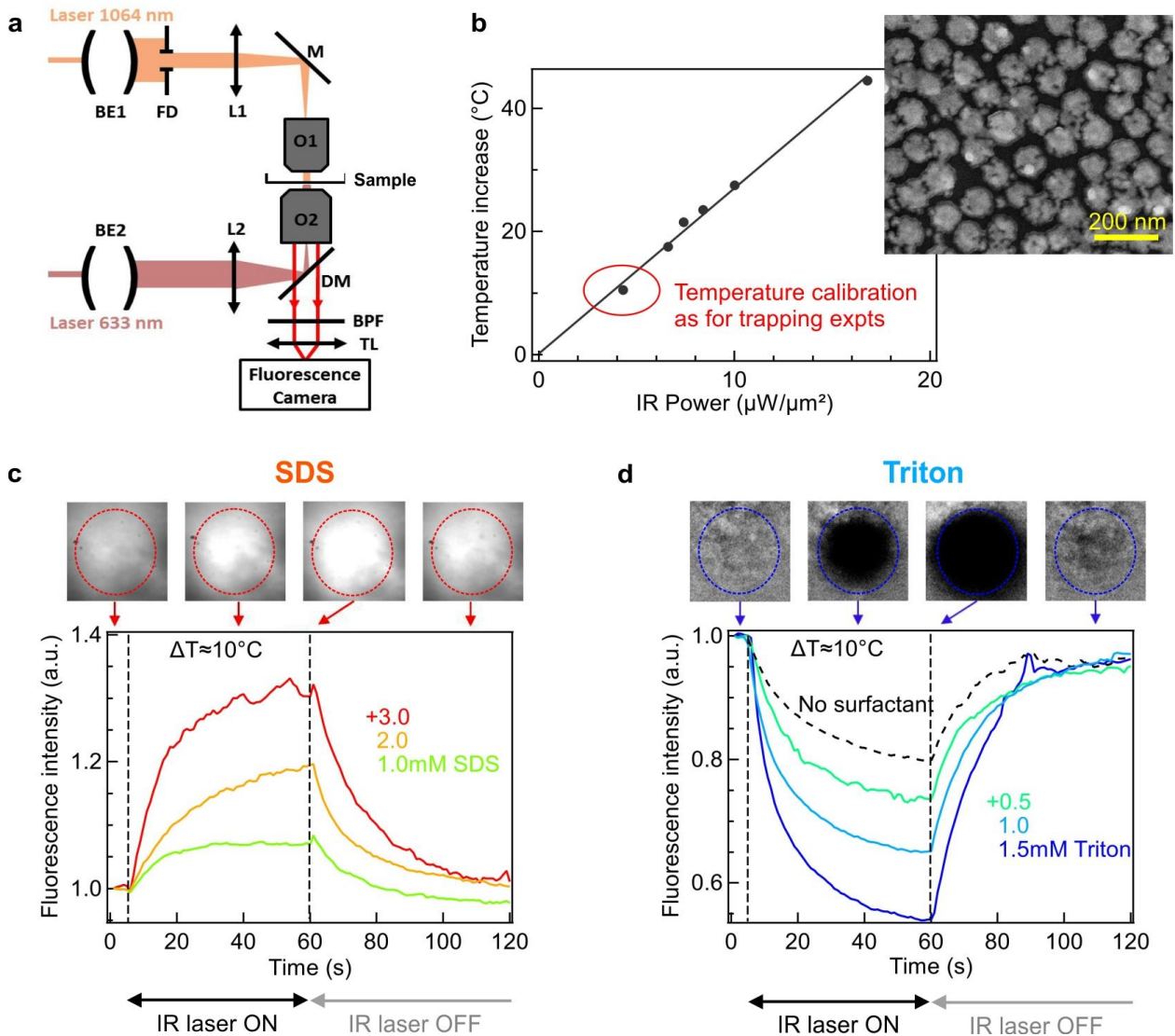

**Figure S9.** (a) Experimental setup for local heating of gold nanoparticle array and simultaneous excitation and fluorescence imaging of 28 nm fluorescent nanoparticles. These polystyrene nanoparticles are the same as in the rest of the study, yet their concentration is increased by 10× to better visualize the concentration gradients. An infrared solid state laser beam is collimated by an air objective (O1, Olympus UPlanFLN 60x, N.A.=0.9) by passing through a beam expander (BE1), field diaphragm (FD), lens (L1) and mirror (M). A red HeNe laser beam is collimated by an oil objective (O2, Olympus UAPON 100x, N.A.=1.49) by passing through a beam expander (BE2), lens (L2) and dichroic mirror (DM). The fluorescence is collected through the same objective (O2), DM, band-pass filter (BPF), tube lens (TL) and imaged on a low noise sCMOS camera (Andor Zyla 5.5). (b) Calibration of the temperature increase performed following our earlier work.[16] The inset shows a SEM image of the gold nanoparticle sample. (c-d) Fluorescence evolutions with time. The IR heating laser starts from 6 to 60s to ensure the local temperature increase around 10°C. Each fluorescence image is recorded per second and 4 images in different regimes adjusted by the contrast are shown in each figure. The circle in the images corresponds to a diameter of 85 µm.

## S8. Relationship between the thermophoretic stiffness and the fluorescence change

Here we show that the thermophoretic stiffness $\kappa_{therm}$ and the fluorescence change ΔF are connected to each other by the Soret coefficient $S_T$ which is defined as the ratio of the thermophoretic mobility $D_T$ by the translational diffusion coefficient for Brownian diffusion $D$: $S_T = D_T/D$.[17,18]

The thermophoretic force is given by $F_{therm} = -k_B T S_T \nabla T$, where $k_B$ is Boltzmann's constant, $T$ the local temperature and $S_T$ the Soret coefficient.[19] So, the thermophoretic stiffness $\kappa_{therm}$ is proportional to $S_T$.

The Soret coefficient also determines the steady-state nanoparticle concentration gradient given by $\nabla C/C = -S_T \nabla T$.[17,18] The fluorescence change ΔF is connected to this nanoparticle concentration change, and hence to the Soret coefficient $S_T$. Therefore, both $\kappa_{therm}$ and ΔF are proportional to $S_T$ and thus to each other.

In the case of the fluorescence change ΔF though, the proportionality is only an approximation valid in the linear regime. Additional effects (triplet blinking, nonradiative rate change due to temperature) may introduce deviations from this linear dependence.[1]

The theoretical thermophoretic stiffness can be estimated using $\kappa_{therm} = k_B T S_T \nabla T/\delta x$. The thermal energy $k_B T$ at 12°C above room temperature amounts to 4.22e-21 J or 4220 fN.nm. The Soret coefficient $S_T$ is estimated to 0.3 K$^{-1}$ from Ref.[20]. Our numerical simulations (Fig. 1d and Ref.[1]) estimate the thermal gradient $\nabla T$ around 10 K/µm at 5 mW/µm² illumination intensity. The lateral displacement around equilibrium $\delta x$ is taken equal to 5 nm. With these reasonable estimates, the theoretically expected thermophoretic stiffness $\kappa_{therm}$ amounts to 0.5 fN/nm/mW. This calculation corresponds very well to the values in Fig. 4b as we find $\kappa_{therm}$ = 0.4 ± 0.1 fN/nm/mW for 1mM SDS and $\kappa_{therm}$ = -0.5 ± 0.1 fN/nm/mW for 1mM Triton.

## S9. Linear-linear plot of the intensity-normalized trap stiffness as a function of ΔF

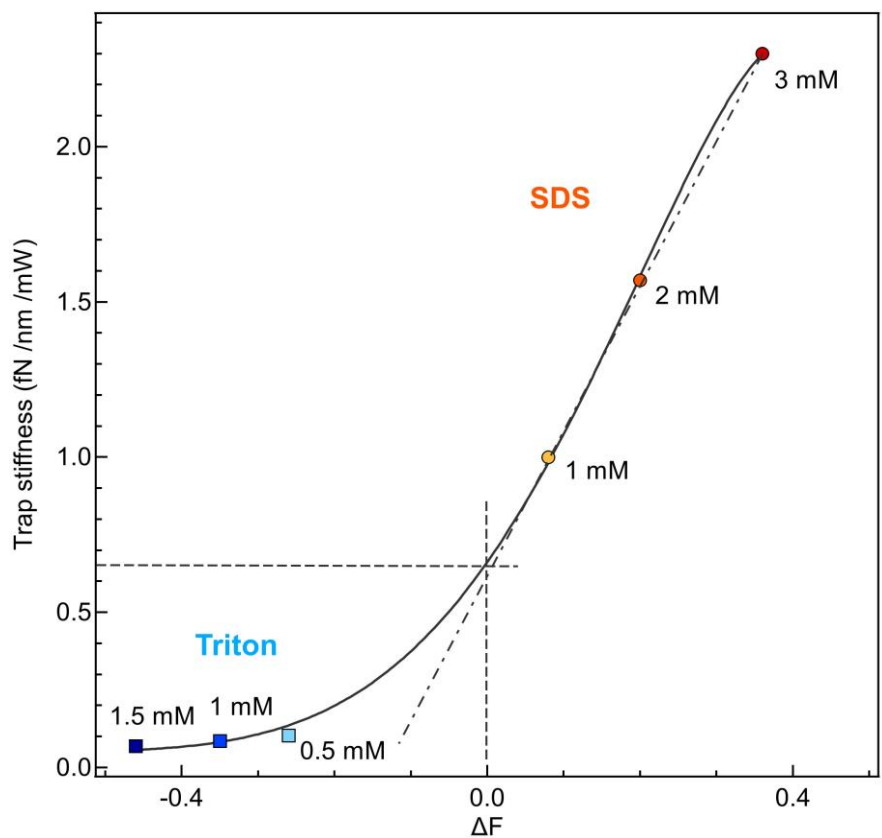

**Figure S10.** Intensity-normalized trap stiffness as a function of ΔF. The numerical interpolation (black curve) is used to determine the trap stiffness $\kappa_{opt}$ at ΔF =0 which we attribute only to the optical force.

**S10. Additional method details**

*DNH fabrication*

Double nanohole apertures are milled on substrates consisting of 150 µm thick borosilicate glass on top of which a 5 nm chromium layer and a 100 nm gold layer are deposited by electron-beam evaporation (Bühler Syrus Pro 710). The DNH structures are then directly milled by focused ion beam (FEI dual beam DB235 Strata) using a gallium ion source with 30 kV voltage and 10 pA beam current.

*Surfactants and nanoparticles*

Sodium dodecyl sulfate and Triton X-100 are purchased from Sigma Aldrich and used directly without further purification. The fluorescent nanoparticles used as objects for the trapping experiments are purchased from Thermofisher Invitrogen (ref F8783). They are carboxylate-modified polystyrene nanospheres with 28 nm diameter (this is the actual diameter measured by the manufacturer and confirmed by our FCS analysis in the Supporting Information Fig. S2). The polystyrene nanoparticles are doped with dark red fluorescent dyes with absorption/emission maxima at 660/680 nm. For the trapping experiments, the fluorescent nanoparticles are diluted to a final concentration of $4.25 \times 10^{11}$ particles/mL. All dilutions are performed in ultrapure water with conductivity 18.2 MΩ.cm (Merck Millipore DirectQ-3 UV). No salts are added to the solution.

*Experimental setup*

We use an inverted confocal microscope with a continuous wave 1064 nm laser (Ventus 1064-2W) for trapping. In addition to this infrared laser, we overlap a 635 nm pulsed laser diode (Picoquant LDH-P-635) at 80 MHz repetition rate and 5 µW average power to excite the fluorescence from the polystyrene nanoparticles. Both lasers are focused by a high NA microscope objective (Zeiss Plan-Neofluar 40x, NA 1.3, oil immersion) to spot diameters of 1 µm and 0.6 µm for the 1064 and 635 nm lasers respectively. The infrared illumination intensity expressed in this paper (in mW/µm²) includes the 50% transmission of the objective at 1064 nm. The fluorescence from the nanoparticles is collected by the same microscope objective in epi- configuration. A set of dichroic mirrors, long pass filters, 30 µm confocal pinhole and bandpass filters ensures that only the fluorescence light is detected, and not the scattered light by any of the two lasers. Two avalanche photodiodes (Picoquant MPD-5CTC) separated by a 50/50 beam-splitter record the fluorescence photons in the 650-690 nm spectral range. The use of two photodiodes avoids the afterpulsing issue in fluorescence correlation spectroscopy (FCS) and make sure there is no artefact in the correlation data even for short lag times below 1µs. The photodiode output is connected to a time correlated single photon counting module (Picoquant Picoharp 300 with PHR 800 router) with time-tagged time-resolved (TTTR) option. All the fluorescence time traces are analyzed with the Symphotime 64 software (Picoquant) enabling to compute the intensity time trace and the intensity relation function. Only trapping events lasting more than 1 s are retained for analysis. We have also checked the ergodicity of our data: similar correlation times were obtained by analyzing a long trapping event than by averaging several correlation times obtained from different shorter trapping events.

*Numerical simulations*

Simulations of the electromagnetic intensity and the temperature distributions around the DNH are performed using COMSOL Multiphysics with the "Electromagnetic Waves, Beam Envelopes" and "Heat Transfer in Solids" modules. The refractive index for gold and chromium are taken from Johnson and Christy, while the thermal parameters for the materials are taken from the COMSOL library.[2] We use the automatic mesh from COMSOL with tetrahedrons of 5 nm side length in the structure. The illumination at 1064 nm is set along a Gaussian distribution with a 1 µm waist and a linear polarization perpendicular to the axis joining the two apertures in the DNH. The reference $E_0$ is taken as the electric field amplitude at the center of the incoming beam.